\documentclass[USenglish,twocolumn]{article}

\usepackage[utf8]{inputenc}%(only for the pdftex engine)
\usepackage[big]{dgruyter}
\usepackage{graphics,graphicx}
 
\def\XMM{{\em XMM--Newton}}
\def\ROSAT{{\em ROSAT}}

\def\Chandra{{\em Chandra}}
\def\Swift{{\em Swift}}

\def\BDone{BD\,+37$^\circ$\,442}   
\def\BDtwo{BD\,+37$^\circ$\,1977}   
\def\BDthree{{BD+28$^\circ$ 4211}}
\def\RX{RX J0648.0--4418}
\def\HDone{HD\,49798}
\def\HDtwo{HD\,160641}
\def\approxgt{\mathrel{\hbox{\rlap{\lower.55ex \hbox {$\sim$}}
        \kern-.3em \raise.4ex \hbox{$>$}}}}
\def\approxlt{\mathrel{\hbox{\rlap{\lower.55ex \hbox {$\sim$}}
        \kern-.3em \raise.4ex \hbox{$<$}}}}
\def\pdot {\dot P}
\def\nudot {\dot \nu}
\def\mdot {\dot M}
\def\flux {\mbox{erg cm$^{-2}$ s$^{-1}$}}
\def\lum {\mbox{erg s$^{-1}$}}

\def\lx {$L_{\rm X}$}
\def\fx {$f_{\rm X}$}
\def\lbol {$L_{\rm BOL}$}
\def\ltsima{$\; \buildrel < \over \sim \;$}
\def\lsim{\lower.5ex\hbox{\ltsima}}
\def\gtsima{$\; \buildrel > \over \sim \;$}
\def\gsim{\lower.5ex\hbox{\gtsima}}
\def\msole{M$_{\odot}$}

\def\lsole{L$_{\odot}$}

\def\countsec{\hbox{cts s$^{-1}$}}

\def\hcm {\hbox {\ifmmode $ atom cm$^{-2}\else atom cm$^{-2}$\fi}}

\begin{document}

  \articletype{Research Article{\hfill}Open Access}

  \author*[1]{Nicola La Palombara}

  \author[2]{Sandro Mereghetti}

  \affil[1]{Affil, E-mail: nicola@iasf-milano.inaf.it}

  \affil[2]{Affil, E-mail: sandro@iasf-milano.inaf.it}

  \title{\huge News on the X-ray emission from hot subdwarf stars}

  \runningtitle{X-ray emission from hot subdwarf stars}

  \begin{abstract}
{In latest years, the high sensitivity of the instruments on-board the \XMM\ and \Chandra\ satellites allowed us to explore the properties of the X-ray emission from hot subdwarf stars. The small but growing sample of X-ray detected hot subdwarfs includes binary systems, in which the X-ray emission is due to wind accretion onto a compact companion (white dwarf or neutron star), as well as isolated sdO stars, in which X-rays are probably due to shock instabilities in the wind. X-ray observations of these low-mass stars provide information which can be useful for our understanding of the weak winds of this type of stars and can lead to the discovery of particularly interesting binary systems. Here we report the most recent results we have recently obtained in this research area.}
\end{abstract}
%  \classification[PACS]{}
 % \communicated{...}
 % \dedication{...}

  \journalname{Open Astronomy}
\DOI{DOI}
  \startpage{1}
  \received{..}
  \revised{..}
  \accepted{..}

  \journalyear{2017}
  \journalvolume{1}
%  \journalissue{1}

\maketitle
\section{Introduction}

In the late seventies the observation of massive OB stars with the first imaging X-ray telescopes identified them as a distinct class of X-ray sources \citep{Seward+79,Harnden+79}. After almost fourty years of studies with telescopes of increasing performances, now it is well known that these stars show X-ray luminosities up to \lx\ $\sim 10^{33}$ \lum\ and spectra of thermal type, with plasma temperatures $T \sim 10^6$ K; moreover, several surveys have shown that the X-ray and bolometric luminosities of these stars are related by the relation \lx/\lbol\ = $10^{-7 \pm 1}$ \citep{Naze09}. These stars are characterized by strong stellar winds, with typical mass-loss rates $\dot M_{\rm W} = 10^{-7} - 10^{-5}$ \msole\ yr$^{-1}$ and terminal velocities of a few thousands km s$^{-1}$. Stellar winds are a common feature of hot stars, irrespective of their wide range of luminosities, masses, and chemical compositions: among the massive stars, strong mass losses have been observed in Wolf--Rayet, O--type, and B--type stars, while among hot low--mass stars stellar winds have been revealed in the central stars of planetary nebulae and in a few extreme helium (EHe) and O--type subdwarf (sdOs) stars \citep{Hamann10}. According to the radiative line-driven wind theory \citep{Castor+75,KudritzkiPuls00}, these winds are accelerated by the photons emitted by the star, which transfer part of their momentum to the wind matter through line absorption and reemission. However, it is well established that these winds are neither steady-state nor homogeneous; they seem to be highly structured on a broad range of spatial scales and several observational results suggest a clumped structure of the stellar winds \citep{Sundqvist+12}. In fact, the non-linear growth of the so-called \textit{line-deshadowing instability} (LDI) leads to high-speed rarefactions that form strong reverse shocks, whereby most wind material is compressed into spatially very narrow clumps \citep{SundqvistOwocki12}. In single, non-magnetic hot stars, the most favoured model is the \textit{embedded wind shock} (EWS) scenario \citep{Owocki13,Owocki+13,Cohen+14}: the X-rays are emitted from hot plasma ($T$ = 1-10 MK), which is heated by the clump–clump collisions and is embedded by the cool wind material.

If the X-ray investigation of massive OB stars is a well recognized research field, the situation is different for the hot subdwarf (sd) stars. Although they are characterized by comparable temperatures and spectral types, their masses and luminosities are significantly smaller than those of massive stars. Therefore, in the Herzprung-Russell (HR) diagram they are well below the main sequence \citep{Heber16}. Historically, they have been deeply investigated in the optical-UV domain, while they remained undetected at X-rays. The only exception is the sdO star \HDone, which was detected already with the \textit{Einstein} observatory in 1979; in this case, however, most of the observed X-ray emission is due not to wind shocks but to the matter accretion onto the compact companion \RX\ \citep{Israel+97}. For more than three decades no other hot subdwarf stars were revealed in this energy range, although it was expected that also this type of stars were characterized by non-negligible stellar winds. In fact, the line-driven wind theory predicts that the mass-loss rate depends on the star luminosity according to the relation $\mdot_{\rm W} \propto L^\alpha$, where $\alpha \sim$ 1.5--2; in the case of hot subdwarf stars this relation implies significant mass-loss rates, although weaker than those of main sequence, giant and supergiant OB stars. These predictions were confirmed for some sdO stars, where the P-Cygni profiles of the C \textsc{iv} and N \textsc{v} lines in their UV spectra imply $\dot M_{\rm W} = 10^{-9} - 10^{-8}$ \msole\ yr$^{-1}$ \citep{Hamann10,JefferyHamann10}. In addition, the extrapolation to lower luminosities of the \lx\ - \lbol\ relation implies \lx\ = $10^{27-32}$ and $10^{26-29}$ \lum\ for O and B type subdwarfs, respectively. Therefore, the detection of intrinsic X-ray emission from hot subdwarf stars was expected.

This result was obtained in the last decade, thanks to the high sensitivity of the \XMM\ and \Chandra\ space telescopes. They allowed us to detect and investigate the X-ray emission of four additional sdO stars: \BDone, \BDtwo, Feige 34, and \BDthree\ \citep{MereghettiLaPalombara16}. \BDone\ and \BDtwo\ are very similar to \HDone, since they have high luminosity (\lbol\ $\sim 10^{4}$ \lsole) and low surface gravity (log $g \sim$ 4); however, both stars are at a distance above 2 kpc, while the estimated distance of \HDone\ is 650 pc \citep{KudritzkiSimon78}. On the contrary, Feige 34 and \BDthree\ are `compact' subdwarf stars, with low luminosity (\lbol\ $< 10^{3}$ \lsole) and high surface gravity (log $g >$ 5); they are also much nearer than the previous stars. For all stars the detected X-ray fluxes are very low (\fx\ $\sim 10^{-14}$ \flux); they imply \lx\ = $10^{30-31}$ and $10^{28-29}$ \lum\ for the `luminous' and `compact' stars, respectively.  Two results suggest that also for these subdwarf stars the observed X-ray emission originates in the stellar wind: in all cases the X-ray/bolometric flux ratio is $\sim 10^{-7}$, in agreement with the relation found for the massive early-type stars; both the spectra of \BDone\ \citep{Mereghetti+17} and \BDtwo\ \citep{LaPalombara+15} were well described with the sum of two plasma emission models with different temperatures, as in the case of the massive stars \citep{Naze09}.

Among these sources, \HDone\ is the only star in a binary system. It is a single-lined spectroscopic binary, with an orbital period $P_{\rm orb} \simeq$ 1.55 d and an optical mass function $f(M)$ = 0.27 \msole\ \citep{Thackeray70,SticklandLloyd94}. The companion star is undetected in the optical-UV band; its nature was revealed in 1992 by \ROSAT, which detected a pulsed X-ray emission with $P$ = 13.2 s \citep{Israel+97}. This short and regular periodicity of the X-ray emission can be explained only with the rotation of a compact companion star (\RX), either a white dwarf (WD) or a neutron star (NS). The observed X-ray emission was attributed to the accretion of part of the stellar wind from the sdO star onto the surface of the compact companion; this is the only known pulsar with a hot subdwarf companion.

The \XMM\ observations performed in the following years \citep{Mereghetti+09,Mereghetti+11b,Mereghetti+13} allowed us to analyse the orbitally-induced phase delays of the X-ray pulses and to measure the X-ray mass function. Moreover, some observations covered also the eclipse phase of the X-ray pulsar, which occurs when \RX\ passes behind \HDone: in this way, from the duration of the X-ray eclipse we derived the system inclination. Together with the two mass functions, this allowed us to obtain a dynamical measurement of the masses of \HDone\ (1.50 $\pm$ 0.05 \msole) and \RX\ (1.28 $\pm$ 0.05 \msole); the latter is consistent with either a NS or a massive WD. The \XMM\ observations provided also a well-constrained measurement of the source flux (\fx\ = $7\times 10^{-13}$ \flux\ in the energy range 0.2-10 keV), which implies an estimated luminosity \lx\ = $3\times 10^{31}$ \lum\ (for $d$ = 0.65 kpc). This low luminosity value suggested that \RX\ is most likely a WD, since a NS would provide a much larger luminosity.

We discovered that during the eclipse phase of \RX\ the X-ray flux decreased by a factor $\sim$ 10 but did not disappear completely, and an X-ray source was still detected. Its flux is \fx\ = $6\times 10^{-14}$ \flux, corresponding to luminosity \lx\ = $2.5\times 10^{30}$ \lum; this implies \lx\ / \lbol\ $\simeq 9\times10^{-8}$. The latter parameter is consistent with the range of values measured for the massive early-type stars, as for the other sdO stars. Moreover, as in the case of \BDone\ and \BDtwo, the spectrum of this component can be well described with the sum of two or three plasma emission models with different temperatures \citep{Mereghetti+16}. Therefore, very probably the X-ray flux of \HDone\ observed when the companion pulsar is eclipsed is produced in the wind of the sdO star itself.

In addition to the five sources described above, we have observed at X-rays other 29 hot subdwarf stars. 16 of them are a complete flux-limited sample of isolated sdO stars, with properties similar to those of the five detected sources. We investigated them in a systematic way using \Chandra, with the aim to search for possible intrinsic X-ray emission \citep{LaPalombara+14}. The remaining 13 sources are nearby sdB stars for which evidence for a close and/or massive compact companion had been reported. Therefore, they were considered good candidates for accreting X-ray sources and observed with either \Swift\ \citep{Mereghetti+11a} or \XMM\ \citep{Mereghetti+14}. None of these stars was detected at X-rays, and we could only set upper limits on their X-ray luminosity.

\section{New results on \HDone: pulsar spin-up and intrinsic X-ray emission}

In the latest years we have exploited all the \XMM\ observations of this source to perform an accurate timing analysis. To this aim, we have considered also the \ROSAT\ PSPC data taken in 1992 and the \Swift\ XRT data taken in the observations performed between 2013 and 2015. We obtained the phases of the pulsations as a function of the time and fitted them with a quadratic function, with the aim to obtain a phase-connected timing solution. We started by considering only the most closely spaced observations (those taken in Summer 2011), while the other observations were gradually included afterwards in the fit. This iterative procedure allowed us to obtain a unique phase-connected solution for all the considered data (over a range of more than 20 years). This analysis revealed that the pulsar is characterized by a spin-up, with a period derivative $\pdot$ = (-2.15 $\pm$ 0.05)$\times 10^{-15}$ s s$^{-1}$ \citep{Mereghetti+16}. This was the first measure of the spin-period derivative for the X-ray pulsar in this system.

From the spectral point of view, we first analysed the 8 \XMM\ observations (performed between 2008 and 2011) which included the eclipse phase. We stacked together all the eclipse data (for a total exposure time of $\sim$ 30 ks), then the resulting spectrum (due to the wind of the sdO star) was fitted with the sum of three thermal plasma components \citep{Mewe+85} at different temperatures and abundances fixed at those of \HDone\ \citep{MereghettiLaPalombara16}. The best-fit model obtained in this way was then considered a fixed component to be included in all the fits of the non-eclipsed X-ray emission.

For the spectral analysis of the X-ray pulsar we summed the out-of-eclipse data of all the \XMM\ observations, which were performed between 2002 and 2014; the corresponding total exposure time was $\sim$ 90 ks. To fit this stacked spectrum, we considered the eclipse model as a fixed component and we fitted an additional model composed by a blackbody plus a power-law component. We found that the pulsar spectrum was dominated by the BB component, with best-fit parameters $R_{\rm BB} = 40 \pm 5$ km and $kT_{\rm BB} = 31 \pm 1$ eV; the corresponding bolometric flux of this component (at 3 $\sigma$ c.l.) is $f_{\rm BB}$ = (3--5)$\times 10^{-12}$ \flux. On the other hand, the PL component provides only a marginal contribution to the total flux, since $f_{\rm PL} \lsim 2\times 10^{-13}$ \flux. Therefore, we estimated that the total accretion-powered luminosity is $L_{\rm accretion}$ = (2.0 $\pm$ 0.5)$\times 10^{32}$ (d/650 pc)$^{2}$ \lum.

The discovery of a long-term spin-up confirms that \RX\ is accreting mass. From the period derivative we can estimate the specific angular momentum which is gained by the compact object: $j = \frac{2 \pi \nudot I G M} {L_{\rm accretion} R}$, where $\nu = 1/P$ is the star rotation frequency, $\mdot$ is the mass accretion rate, and $M$, $R$, and $I$ are the mass, radius and moment of inertia of the compact object, respectively. The specific angular momentum is very different for the WD and the NS cases, since $j_{WD} = 2.2\times10^{19} \left({L_{\rm accretion}}\over{2\times10^{32}~{\rm erg~s^{-1}}}\right)^{-1} ~{\rm cm^2~s^{-1}}$ in the case of a WD (for which we assume $I = 10^{50}$ g cm$^2$, $R = 3000$ km, $M$ = 1.28 \msole), while $j_{NS} = 5.5\times10^{16} \left({L_{\rm accretion}}\over{2\times10^{32}~{\rm erg~s^{-1}}}\right)^{-1} ~{\rm cm^2~s^{-1}}$ in the case of a NS (where $I = 10^{45}$ g cm$^2$, $R = 12$ km, $M$ = 1.28 \msole). The values obtained in this way should be compared with the specific angular momentum that can be provided by the inflow of the accreting matter.

In the case of wind accretion, this parameter depends strongly on the relative velocity between the compact object and the stellar wind ($V_{\rm REL}$). For \HDone\ we estimated that $j_w = 5.4\times10^{16} \left({V_{\rm REL}}\over{1000 ~{\rm km~s^{-1}}}\right)^{-4} ~{\rm cm^2~s^{-1}}$. This implies that it would be difficult to obtain the observed spin-up rate if the compact object is a WD. On the other hand, the wind-accretion scenario seems more likely if the compact object is a NS. However, in this case the estimated angular momentum would require a rather small value of $V_{\rm REL}$; moreover, since for a NS $L_{\rm NS} = 1.3\times10^{34}~\left({\mdot} \over {{\rm 10^{17}~g~s^{-1}}} \right) \left( {V_{\rm REL}} \over {{\rm 1000~{\rm km~s^{-1}}}} \right)^{-4} ~~{\rm erg~s^{-1}}$, this low value of $V_{\rm REL}$ would imply a luminosity much higher than the observed one.

Due to these difficulties, we considered the alternative scenario of disc accretion. Taken into account the minimum magnetic field required to provide the observed spin-up rate and the constraints on the magnetospheric and corotation radii, we obtained a lower limit on the accretion rate. For the WD case $\mdot > 2\times 10^{16}$ g s$^{-1}$. This limit is rather high and implies a high luminosity, which can be in agreement with the observed flux only if the source distance is larger than 4 kpc, while the commonly adopted distance in the literature is 650 pc \citep{KudritzkiSimon78}. In the case of a NS, the limit on the accretion rate is much lower ($\mdot > 2\times 10^{11}$ g s$^{-1}$) and implies a luminosity which agrees with the observed one. But such a low accretion rate requires a very low magnetic field ($2 \times 10^7$ G $< B_{\rm NS} < 3 \times 10^{10}$ G), which is unusual in accreting pulsars in high mass X-ray binaries (although the central compact objects of some supernovae remnants are young NSs with $B_{\rm NS}$ consistent with the above range); moreover, the stability of the spin-up rate, maintained for a time period longer than 20 yr, is very difficult to explain; finally, the large radius of the emitting area derived from the blackbody spectral fit ($R_{\rm BB} \simeq$ 40($d$/650 pc) km) is inconsistent with the NS size.

In summary, in the latest two years we have obtained the first measurement of the spin-period derivative for \RX, the X-ray pulsar in the binary system of \HDone. We have found that it is characterized by a high spin-up rate. It is rather large for a WD but, due to the uncertainties in the distance measurement, we cannot reject this possibility; in any case, a WD interpretation would require the presence of an accretion disc. On the other hand, a NS interpretation would indicate a binary with peculiar properties, very different from those of all the other known NS X-ray binaries. The uncertainty about the pulsar nature could be resolved with a reliable estimate of the source distance. Therefore, the parallax measurement obtained with \textit{Gaia}, to be released in 2018, will be an essential parameter.

\section{New results on \BDone: lack of X-ray pulsations and stellar wind emission}

\BDone\ was the second subdwarf star detected at X-rays \citep{LaPalombara+12}. It is a very luminous (\lbol\ $\sim 10^{38}$ \lum) and He-rich sdO star \citep{Rebeirot66,Husfeld87}. Its temperature, luminosity, surface gravity, and mass-loss properties are very similar to those of \HDone. Contrary to this star, however, for \BDone\ there was no evidence for a binary companion \citep{FayHoneycuttWarren73,KaufmannTheil80,DworetskyWhitelockCarnochan82}; also recently high-resolution spectroscopical observations did not reveal any radial velocity variations \citep{Heber+14}. In 2011 we observed this star with \XMM\ and detected a soft X-ray emission. The source spectrum was described with a PL+BB model with $kT_{\rm BB} = 45 \pm 10$ eV and $\Gamma_{\rm PL} \sim 2$, thus similar to that of \HDone\ out of the eclipse phase. The source flux was \fx\ = $3\times 10^{-14}$ \flux\ but, due to the large uncertainties on the spectral parameters, the estimated luminosity was poorly constrained (\lx\ $\sim 10^{32} - 10^{35}$ \lum). We also found a periodic modulation of the X-ray flux, with $P$ = 19.156 $\pm$ 0.001 s, at a 3.2 $\sigma$ statistical significance. This unexpected result suggested the possible presence of a compact binary companion, but it was at odds with the absence of variations in the radial velocity. This contradiction can be explained if the orbital plane has a very small inclination and/or the orbital period is very long (of the order of several months). The first possibility seems unlikely, because the large projected rotational velocity ($\sim$ 60 km s$^{-1}$) observed by \citet{Heber+14} would imply a significant misalignment between the star rotation axis and the orbital axis. The second possibility would imply a low mass accretion rate, unless the accreting companion is observed close to the periastron passage in a very eccentric orbit.

In order to investigate more deeply the different possibilities, in 2016 we performed a new \XMM\ observation, longer than the previous one (50 ks instead of 30). This time we found no evidence of periodicity in the X-ray flux, even by exploring the range of periods between 19.1 and 19.2 s to take into account a possible spin-up or spin-down of the source since the previous observation \citep{Mereghetti+17}. We performed Monte Carlo simulations to verify that a pulsation like that observed in 2011 should have been detected with a high confidence level also in the new observation. This result suggests that either the pulsations in this source disappeared (since they became undetectable due to a decrease of the pulsed fraction) or that the peak at 19.2 s detected in 2011 in the power distribution was caused by a statistical fluctuation.

The spectral analysis of the new data showed that, compared to the previous observation, there was no significant difference in the spectral shape and in the source flux. Therefore, we performed a simultaneous spectral analysis of the two spectra. We described the observed spectra with a single-temperature plasma model (\textsc{apec} in \textsc{xspec}), finding strong evidence for an overabundance of C and Ne. Since \BDone\ is an extreme He-rich star, we considered two different solutions for the element abundances: either the He abundance fixed to a mass fraction X$_{\rm He}$ = 0.99, the C and Ne abundances free to vary, and the abundances of the other elements fixed to the solar values; or the abundances of He, C, N, Si and Fe fixed at the values of \citet{JefferyHamann10} and only the Ne abundance free to vary. In both cases we obtained an equally good fit. With these models the measured flux was in the range \fx\ = (3.3--6.6)$\times 10^{-14}$ \flux, corresponding to luminosities \lx\ $\sim$ (1.5--3.0)$\times 10^{31}$ \lum\ for a distance of 2 kpc \citep{BauerHusfeld95}.

The results of both the timing and the spectral analysis of the new \XMM\ observation have changed our view of \BDone: on one hand, in the new X-ray data we could not detect any significant pulsation; on the other hand, we already knew that there are no variations in the radial velocity. Both these findings strongly suggest that \BDone\ is a single star without any companion and that the observed X-ray emission originates from the sdO star itself. This hypothesis is supported by other data: the estimated X-ray luminosity implies an X-ray-to-bolometric ratio \lx\ / \lbol\ $\sim 2\times10^{-7}$, a value consistent with the average relation observed in normal O-type stars; we already know from UV and optical spectroscopy that this star is characterized by a significant stellar wind \citep{JefferyHamann10}; finally, the estimated plasma temperature is comparable to the lowest values seen in the sample of O-type stars observed with \XMM\ \citep{Naze09}. Therefore, we conclude that the observed X-ray emission is most probably related to the shock-heated plasma produced by instabilities in the radiation-driven stellar wind of the sdO star.

\section{The extreme He star \HDtwo}

\begin{figure}[]
\centering
\resizebox{\hsize}{!}{\includegraphics[width=8.5cm,angle=0]{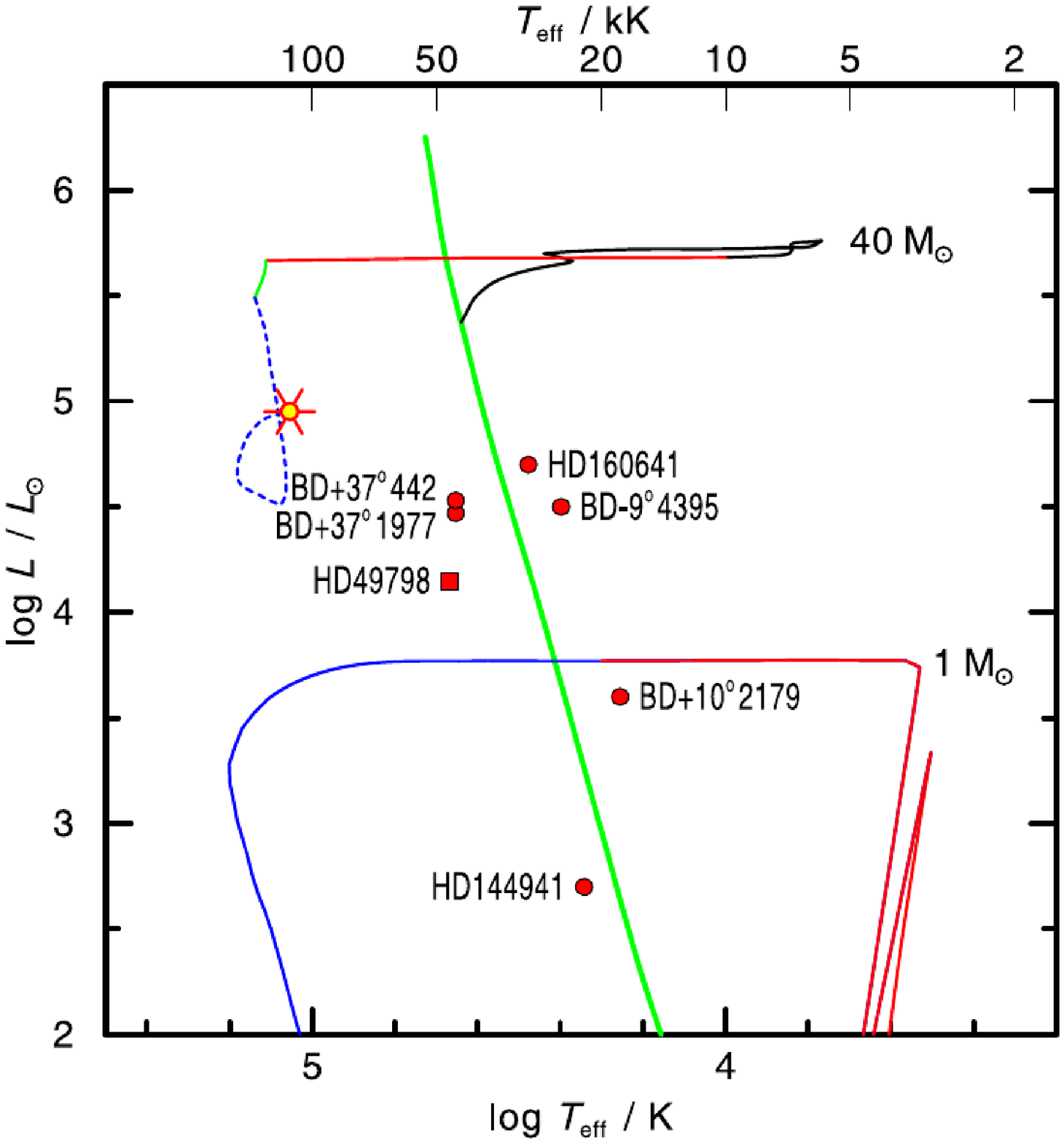}}
\resizebox{\hsize}{!}{\includegraphics[width=8.5cm,angle=0]{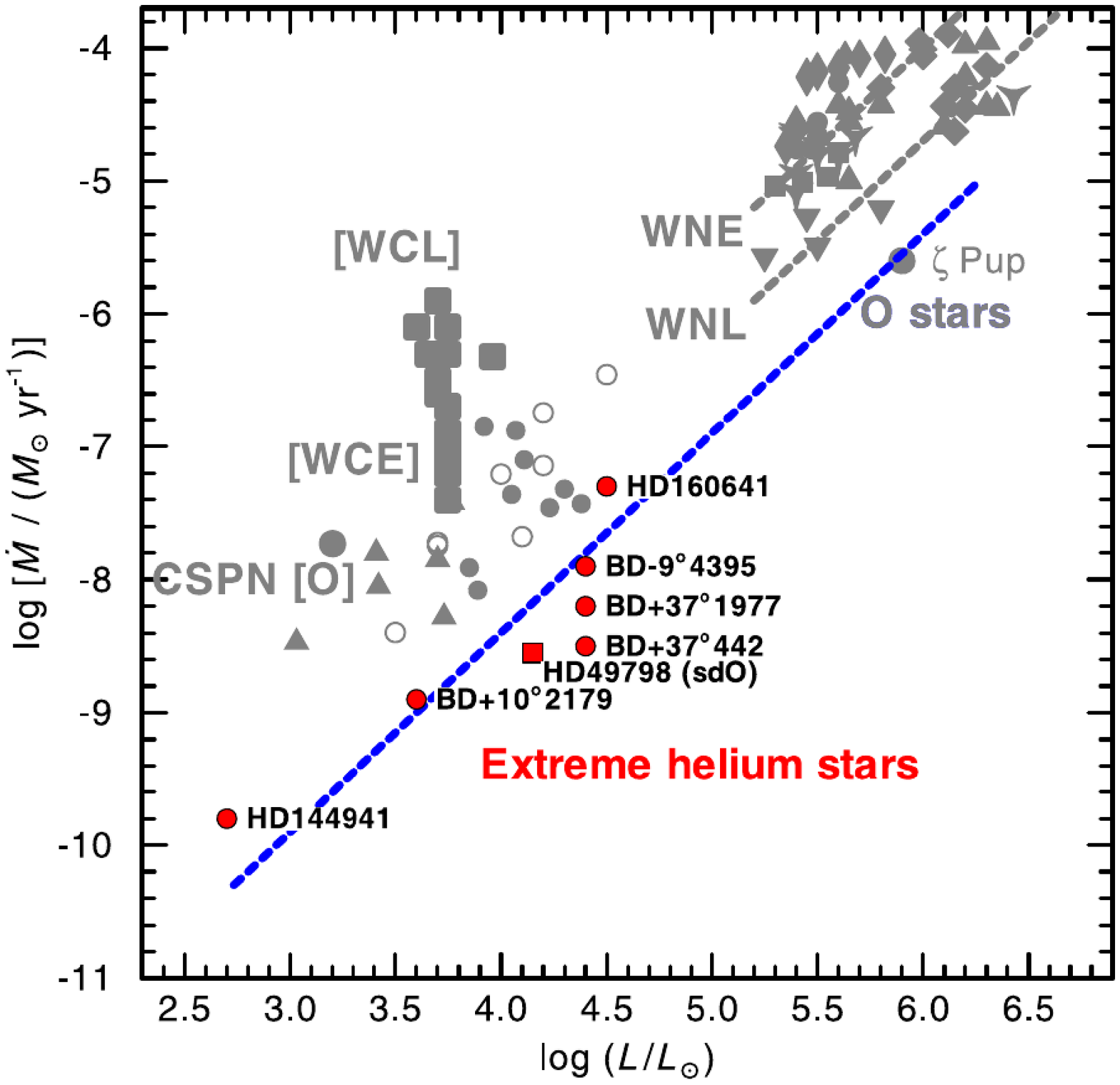}}
\caption{\footnotesize{Position of the sdO stars \HDone, \BDone, and \BDtwo\ in the temperature--luminosity (\textit{top}) and luminosity--mass loss (\textit{bottom}) diagram (from \citet{Hamann10}).}}
\label{TLMdot}
\end{figure}

An interesting class of hot stars is constituted by the extreme helium stars. They are very rare low--mass supergiants in a late stage of evolution. The surfaces of these luminous stars are almost completely depleted from hydrogen: they are composed essentially of helium, with some traces of carbon, and the hottest of these EHe stars are He--rich sdOs stars. Using the latest generation of models for spherically expanding stellar atmospheres, \citet{JefferyHamann10} managed to estimate the mass--loss rates for six of these stars (Table~\ref{EHe_stars}); in Fig.~\ref{TLMdot} we report their position in the HR and in the luminosity/mass--loss rate diagrams. Although the luminosities and mass--loss rates of these stars are lower than those of normal O--type stars, they suggest the possibility that even EHe stars might have intrinsic X--ray emission. This was already demonstrated for the first two stars, \BDone\ \citep{LaPalombara+12,Mereghetti+17} and \BDtwo\ \citep{LaPalombara+15}, which are also those with the highest temperature and surface gravity. Therefore, we aimed to extend our research also to EHe stars which are characterized by lower temperatures and gravities.

\begin{table}[htbp]
\caption{\footnotesize{Parameters of EHe stars and of the He--rich sdO star \HDone\ (from \citet{Hamann10}).}}\label{EHe_stars}
\vspace{-0.5cm}
\begin{scriptsize}
\begin{center}
\begin{tabular}{lccccc} \hline \hline
Source			& T$_{\rm eff}$	& log$g$	& log$L$	& log$\dot M$		& V	\\
			& (kK)		& (cm s$^{-2}$)	& (L$_{\odot}$)	& (M$_{\odot}$/yr)	& (mag)	\\ \hline
\BDone\			& 48		& 4.0		& 4.4		& -8.2			& 10.01	\\
\BDtwo\			& 48		& 4.0		& 4.4		& -8.5			& 10.17	\\ \hline
HD160641		& 35.5		& 2.7		& 4.5		& -7.3			& 9.85	\\
BD-9$^\circ$ 4395	& 25.1		& 2.5		& 4.4		& -7.9			& 10.50	\\
BD+10$^\circ$ 2179	& 18.5		& 2.6		& 3.6		& -8.9			& 9.93	\\
HD144941		& 27.0		& 3.9		& 2.7		& -9.8			& 10.02	\\ \hline
\HDone\			& 46.5		& 4.35		& 4.15 		& -8.5			& 8.29	\\ \hline
\end{tabular}
\end{center}
\end{scriptsize}
\vspace{-0.5cm}
\end{table}

\begin{figure}[t!]
\centering
\resizebox{\hsize}{!}{\includegraphics[width=8.5cm,angle=-90]{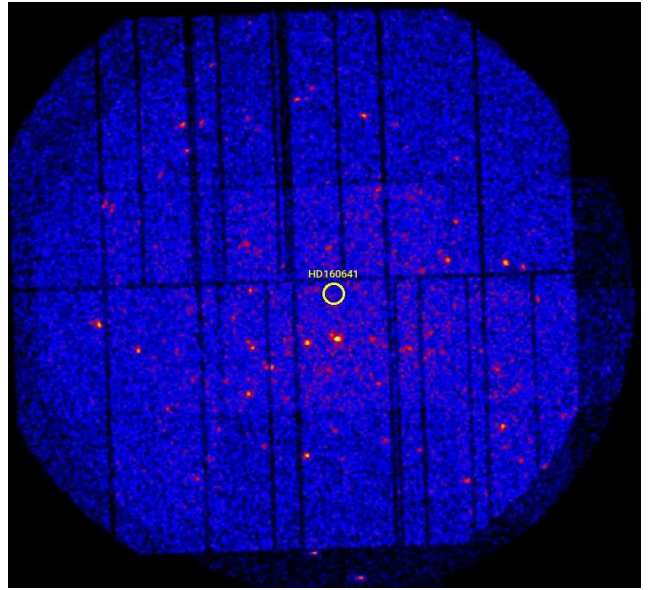}}
\caption{\footnotesize{Image of the \XMM\ field of view around the position of \HDtwo}}
\label{HD160641}
\end{figure}

Among these stars, \HDtwo\ is the best candidate. In fact, as shown in Fig.~\ref{TLMdot}, it has the highest luminosity and mass-loss rate and it is very near to the two BD stars. Moreover, its magnitude is comparable to those of \BDone\ and \BDtwo. Therefore, if it is characterized by a comparable X--ray/optical flux ratio, we expected to detect it at a similar flux level. In September 2016 we performed a 50 ks observation of this source with \XMM, which provided no detection of this star. We searched for a possible X-ray emission in the whole \XMM\ energy range (0.2-12 keV), in order to maximize the count statistics, and in several narrow sub-ranges, to take into account the possibility that the source emission concentrates in restricted energy bands. In all cases we found no significant X-ray emission corresponding to \HDtwo. In Fig.~\ref{HD160641} we report an image of the sky region observed with \XMM, which clearly shows the absence of any X-ray source at the position of our target. We could set an upper limit on the source count rate in the whole energy range: $ CR < 1.5 \times 10^{-3}$ \countsec\ (at 3 $\sigma$ c.l.). If we assume the same thermal-plasma emission model used for the detected EHe stars, from this result we can estimate an upper limit \fx\ $\lsim 10^{-15}$ \flux\ on the source flux, which implies \lx\ $\lsim 6\times10^{29}$ \lum\ for a source distance $d \sim$ 2.3 kpc \citep{JefferyHamann10}. Since for this star \lbol\ $\simeq 1.3\times10^{38}$ \lum, the corresponding upper limit on the X-ray/bolometric flux ratio is \lx\ / \lbol\ $\lsim 5\times10^{-9}$, a value much lower than the average relation observed in the detected sdO stars. Therefore, it is possible that the wind properties of \HDtwo\ are significantly different from those of these stars: although this star has high mass-loss rate and bolometric luminosity, its wind could be much more homogeneous than in the typical O and sdO stars and, then, the internal shocks due to clump collisions could be much less frequent and powerful, thus hampering the plasma heating and the X-ray emission.

\section{Summary and future perspectives}

\begin{figure}[t!]
\centering
%\resizebox{\hsize}{!}{\includegraphics[angle=-90,width=8.5cm]{HD_lc_eclisse.ps}}
\resizebox{\hsize}{!}{\includegraphics[angle=0,width=8.5cm,trim={1.5cm 1cm 1cm 1.5cm},clip=true]{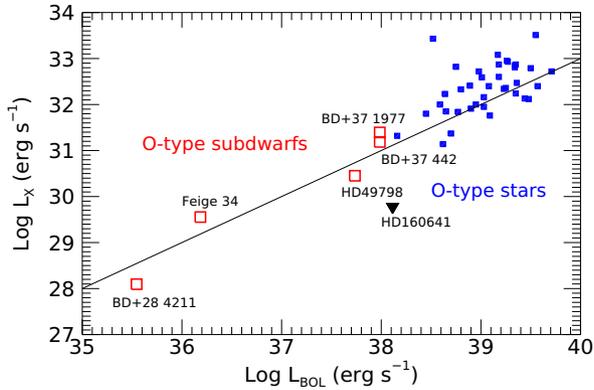}}
\vspace{-0.5cm}
\caption{\footnotesize{X-ray versus bolometric luminosity for O-type subdwarfs (open red squares) and normal O-type stars (small blue squares); for the undetected EHe star \HDtwo\ the luminosity upper limit is reported (black triangle). The line indicates the relation $L_{\rm X} / L_{\rm bol} = 10^{-7}$.}}
\label{lxlbol}
\vspace{-0.5cm}
\end{figure}

Up to now we have observed 21 sdO stars, 13 sdB stars in binary system with a possible compact companion, and 1 EHe star. Only 5 sdO stars were clearly detected, while for all the other we could obtain only an upper limit on the source flux. We know from optical spectroscopy that both the sdO stars \HDone, \BDone, and \BDtwo, and the EHe star \HDtwo\ are characterized by a significant stellar wind; moreover, we expect that a similar wind is present also in the other sdO stars. Therefore, it is interesting to compare their estimated X-ray luminosity (or its upper limit, in the case of the undetected stars) with the bolometric one. To this aim, in Fig.~\ref{lxlbol} we report \lx\ as a function of \lbol\ (in the case of \HDone\ we plot the value corresponding to the eclipse phase of the companion pulsar, which can be attributed to the sdO star itself); for comparison, we report the same parameters also for a sample of normal O-type stars seen in the \ROSAT\ All Sky survey \citep{Berghoefer+96}. We note that the detected sdO stars lie close to the extrapolation of the average X-ray-to-bolometric relation observed at higher luminosities, which suggests that also in the sdO case the observed X-rays are emitted by the shock-heated gas in the stellar winds. Also in the case of the undetected stars, the upper limits are consistent with this relation; hence, we cannot reject the possibility that the same type of emission is present also in these stars, and that their missing detection is due only to the low sensitivity of our observations. The EHe star \HDtwo\ is well below the average relation. Therefore, for this star the intrinsic X-ray emission must be significantly weaker.

The results obtained so far demonstrate the discovery potential of the X-ray observations of hot subdwarf stars. They can be an important tool to probe the radiation-driven winds in these stars, which have a lower mass-loss rate than normal OB stars; moreover, they can be used to detect and study possible accreting compact companions, thus providing useful hints from the evolutionary point of view. Therefore, it is necessary to further investigate in this direction.

Looking at the next future, our first aim is to perform a deep follow-up observation with \XMM\ also for Feige 34 and \BDthree, the other two sdO stars detected up to now. This is particularly important since, while the other detected sdO stars are `luminous' and He-rich sdO stars with a low surface gravity (log($g$) = 4), these two sources are `compact' He-poor stars with high surface gravity (log($g$) $>$ 6). Moreover, while the three sdOs already observed by \XMM\ are among the few hot subdwarfs for which evidence of mass loss has been reported \citep{JefferyHamann10}, there is no comparable evidence in the case of Feige 34 and \BDthree\ \citep{Latour+13}. They are the first stars of this type detected at X-rays, and it would be very interesting to investigate them in detail: the \XMM\ observation of these stars would allow us to investigate the mass loss through radiatively driven winds in a range of surface gravity which has been unexplored so far.

With the next release of the parallax measurements performed with \textit{Gaia} it will be possible to obtain precise distances for several stars: in this way, we will constrain the source luminosities and assess the nature of the compact companion of \HDone. We wish also to increase the sample of subdwarf stars detected at X-rays, in order to explore the possible impact of different parameters (such as surface gravity, temperature and composition) and to further investigate the comparison with normal early-type stars. Only a very small number of hot subdwarf stars have been observed so far, so even with the current telescopes we could obtain many new detections. 

However, currently for the faint sources it is not possible to perform an accurate spectral analysis such as in the bright O-type stars: due to the limited count statistics, in our spectral analysis we have to fix the element abundances to the values provided by the optical studies (when available), since they can not be constrained with the X-ray data alone. From this point of view, a leap forward will be provided by \textit{Athena}, the next large X-ray satellite approved by the European Space Agency. Thanks to its high sensitivity and spectral resolution, it will be possible to accurately measure the plasma temperatures and to constrain within 10 \% the element abundances, in an independent way from optical/UV analyses.

{} 

\end{document}